\documentclass[prl,twocolumn,showpacs,superscriptaddress,amsfonts,amsmath,floatfix]{revtex4}

\usepackage{graphicx}

\begin{document}

\title{Fermi-Bose transformation for the time-dependent Lieb-Liniger gas}

\author{H.~Buljan}
\email{hbuljan@phy.hr (correspondence)}
\affiliation{Department of Physics, University of Zagreb, PP 332, Zagreb, Croatia}
\author{R.~Pezer}
\affiliation{Faculty of Metallurgy, University of Zagreb,
44103 Sisak, Croatia}
\author{T.~Gasenzer}
\affiliation{Institut f\" ur Theoretische Physik, 
Universit\" at Heidelberg, Philosophenweg 16, 69120 Heidelberg, Germany}

\date{\today}

\begin{abstract}
Exact solutions of the Schr\"odinger equation describing a freely 
expanding Lieb-Liniger (LL) gas of delta-interacting bosons in one spatial dimension are constructed.
The many-body wave function is obtained by transforming a fully 
antisymmetric (fermionic) time-dependent wave function which obeys the 
Schr\" odinger equation for a free gas. This transformation 
employs a differential Fermi-Bose mapping operator which depends on the 
strength of the interaction and the number of particles. 

\hfill HD--THEP--07--26\\[-5ex]
\end{abstract}

\maketitle

Nonequilibrium phenomena in quantum many-body systems are among the 
most fundamental and intriguing phenomena in physics. One-dimensional 
(1D) interacting Bose gases provide a unique opportunity to study 
such phenomena. In some cases, the models describing these systems 
\cite{Lieb1963,Girardeau1960,McGuire1964} allow to determine exact 
time-dependent solutions of the Schr\"odinger equation
\cite{Girardeau2000,Girardeau2003} providing insight beyond various 
approximations, which is particularly important in strongly correlated regimes. 
These 1D systems are experimentally realized with atoms tightly 
confined in effectively 1D waveguides \cite{OneD,TG2004,Kinoshita2006}, 
where nonequilibrium dynamics is considerably affected by 
the kinematic restrictions of the geometry \cite{Kinoshita2006},
while quantum effects are enhanced \cite{Olshanii,Petrov,Dunjko}. 
Today, experiments have the possibility to explore 1D Bose 
gases for various interaction strengths, from the Lieb-Liniger (LL) gas 
with finite coupling \cite{OneD,Kinoshita2006} up to the so-called 
Tonks-Girardeau (TG) regime of "impenetrable-core" bosons 
\cite{TG2004,Kinoshita2006}. 
However, most theoretical studies of the exact time-dependence 
address the TG regime (see, e.g., Refs.~\cite{Girardeau2000,Girardeau2000a,
Ohberg2002,Rigol2005,Minguzzi2005,DelCampo2006,Rigol2006,Pezer2007}). 
In this limit, the complex many-body problem is considerably simplified 
due to the Fermi-Bose mapping property \cite{Girardeau2000} 
where dynamics follows a set of uncoupled single-particle 
(SP) Schr\" odinger equations \cite{Girardeau2000}. 
It is therefore desirable to employ an efficient method for calculating 
the time-evolution of a LL gas with finite interaction strength.

In 1963, Lieb and Liniger \cite{Lieb1963} presented, on the basis of the Bethe ansatz, a solution for 
a homogeneous Bose gas with (repulsive) $\delta$-function interactions, 
for arbitrary interaction strength $c$; 
periodic boundary conditions were imposed. 
This system was analyzed by McGuire on an infinite line 
with attractive interactions \cite{McGuire1964}. The renewed 
interest in 1D Bose gases stimulated recent studies 
of static LL wave functions \cite{Muga1998,Sakmann2005,Batchelor2005} including 
a LL gas in box confinement \cite{Batchelor2005}. 
Besides the wave functions, the correlations of a LL system 
with finite coupling 
\cite{Creamer1981,Jimbo1981,Korepin1993,Kojima1997,Olshanii2003,Gangardt2003,
Astrakharchik2003,Forrester2006,Caux2007,Calabrese2007} provide a link 
to many observables 
and were analyzed by using various techniques, including
the inverse scattering method \cite{Korepin1993,Kojima1997,Caux2007,Calabrese2007}, 
$1/c$ expansions \cite{Jimbo1981} relying 
on the analytic results in the TG regime \cite{Lenard1964},
and numerical Quantum Monte Carlo techniques \cite{Astrakharchik2003}.
Regarding dynamics, a full numerical study of the 
irregular dynamics in a mesoscopic LL system was presented in \cite{Berman2004}. 
In Ref. \cite{Girardeau2003}, Girardeau has shown that phase imprinting 
by light pulses conserves the so-called cusp condition 
imposed by the interactions on the LL wave functions, 
and suggested to use time-evolving SP wave functions to analyze 
the subsequent dynamics.
However, as pointed out in Ref. \cite{Girardeau2003}, the presented scheme does not obey the
cusp condition during the evolution which limits its validity. 
This situation can be remedied by using an ansatz 
which obeys the cusp condition at all times by construction \cite{Gaudin1983,Korepin1993}.

Here we construct exact solutions for the freely expanding LL gas
with localized initial density distribution. 
This can be achieved by differentiating a fully antisymmetric 
(fermionic) time-dependent wave function, which obeys the Schr\" odinger 
equation for a free Fermi gas \cite{Gaudin1983}; 
the employed differential operator depends on the interaction 
strength $c$ and the number of particles. 
When $c\rightarrow \infty$, the scheme reduces to Girardeau's 
time-dependent Fermi-Bose mapping \cite{Girardeau2000}, 
valid for "impenetrable-core" bosons.

We consider the dynamics of $N$ indistinguishable $\delta$-interacting bosons 
in a 1D geometry \cite{Lieb1963}.
The Schr\" odinger equation for this system is 
\begin{equation}
i \frac{\partial \psi_B}{\partial t}=
-\sum_{i=1}^{N}\frac{\partial^2 \psi_B}{\partial x_i^2}+
\sum_{1\leq i < j \leq N} 2c\,\delta(x_i-x_j)\psi_B, 
\label{LLmodel}
\end{equation}
where $\psi_B(x_1,\ldots,x_N,t)$ is the many-body wave function, 
and $c$ quantifies the strength of the interaction 
(for connection to physical units see, e.g., \cite{Girardeau2003}).
The $x$-space is infinite (we do not impose any boundary conditions), which 
corresponds to a number of interesting experimental situations 
where the gas is initially localized within a certain region of 
space and then allowed to freely evolve. This is relevant for 
free expansion \cite{Ohberg2002,Rigol2005,Minguzzi2005,DelCampo2006} or 
interference of two initially separated clouds during 
such expansion \cite{Girardeau2000a}, etc. 
Due to the Bose symmetry, 
it is sufficient to express the wave function $\psi_B$ in a single 
permutation sector of the configuration space, $R_1:x_1<x_2<\ldots<x_N$. 
Within $R_1$, $\psi_B$ obeys 
\begin{equation}
i\partial \psi_B/\partial t = 
-\sum_{i=1}^{N} \partial^2 \psi_B/ \partial x_i^2,
\label{free}
\end{equation} 
while interactions impose boundary conditions 
at the borders of $R_1$ \cite{Lieb1963}:
\begin{equation}
\left [
1-\frac{1}{c}
\left ( 
\frac{\partial}{\partial x_{j+1}}-\frac{\partial}{\partial x_j}
\right)
\right]_{x_{j+1}=x_j}\psi_B=0.
\label{interactions}
\end{equation}
This constraint creates a cusp in the many-body wave function 
when two particles touch, which should be present at 
any time during the dynamics.

In the TG limit (i.e., when $c\rightarrow \infty$) the cusp 
condition is $\psi_B(x_1,\ldots,x_j,x_{j+1},\ldots,x_N,t)|_{x_{j+1}=x_j}=0$ 
\cite{Girardeau1960,Girardeau2000},
which is trivially satisfied by an antisymmetric fermionic wave function 
$\psi_F(x_1,\ldots,x_N,t)$; 
thus $\psi_B=\psi_F$ within $R_1$, which is the famous Fermi-Bose 
mapping \cite{Girardeau1960,Girardeau2000}. In many physically interesting 
cases, $\psi_F$ can be constructed as a Slater determinant 
\begin{equation}
\psi_F(x_1,\ldots,x_N,t)=(N!)^{-\frac{1}{2}}
\det[\phi_m(x_j,t)]_{m,j=1}^{N}.
\label{Slater}
\end{equation}
Since $\psi_B=\psi_F$ within $R_1$, 
$\psi_F$ must obey $i\partial \psi_F /\partial t= 
-\sum_{j=1}^{N} \partial^2 \psi_F/ \partial x_j^2$,
which implies that the (orthonormal) SP 
wave functions $\phi_m(x_j,t)$ evolve according to  
\begin{equation}
i \partial \phi_m / \partial t=
 - \partial^2 \phi_m / \partial x^2;
\label{master}
\end{equation}
$m=1,\ldots,N$. 
Thus, in the TG limit, the complexity of the many-body dynamics 
is reduced to solving a simple set of uncoupled SP 
equations, while the interaction constraint (\ref{interactions}) 
is satisfied by the Fermi-Bose construction. 

The simplicity and success of this idea motivates us 
to choose an ansatz which automatically 
satisfies constraint (\ref{interactions}) for any finite $c$ 
\cite{Gaudin1983,Korepin1993}. For this, 
define a differential operator 
\begin{equation}
\hat O=\prod_{1\leq i < j \leq N} \hat B_{ij},
\label{oO}
\end{equation}
where $\hat B_{ij}$ stands for 
\begin{equation}
\hat B_{ij}=\left[
1+\frac{1}{c}
\left(
\frac{\partial}{\partial x_{j}}-
\frac{\partial}{\partial x_{i}}
\right)
\right]
\label{Bij}.
\end{equation}
It can be shown that the wave function 
\begin{equation}
\psi_{B}= {\mathcal N}_{c} \hat O \psi_F
\mbox{ (inside $R_1$)},
\label{ansatz}
\end{equation}
where ${\mathcal N}_{c}$ is a normalization constant, 
obeys the cusp condition (\ref{interactions}) by construction \cite{Gaudin1983,Korepin1993}: 
Consider an auxiliary wave function 
\begin{eqnarray}
\psi_\mathrm{AUX}(x_1,\ldots,x_N,t) & = & \hat B_{j+1,j} \hat O \psi_F \nonumber \\
& = & \hat B_{j+1,j} \hat B_{j,j+1}\hat O'_{j,j+1} \psi_F,
\end{eqnarray}
where the primed operator $\hat O'_{j,j+1}=\hat O/\hat B_{j,j+1}$ omits the 
factor $\hat B_{j,j+1}$ as compared to $\hat O$. 
The auxiliary function can be written as 
\begin{equation}
\psi_\mathrm{AUX}= \left [
1-\frac{1}{c^2}
\left ( 
\frac{\partial}{\partial x_{j+1}}-\frac{\partial}{\partial x_j}
\right)^2
\right] \hat O'_{j,j+1} \psi_F.
\label{AuxAsym}
\end{equation}
It is straightforward to verify that the operator 
$\hat B_{j+1,j} \hat B_{j,j+1}\hat O'_{j,j+1}$ in front of $\psi_F$ is 
invariant under the exchange of $x_j$ and $x_{j+1}$ \cite{Korepin1993}. 
On the other hand, the fermionic wave function $\psi_F$ is 
antisymmetric with respect to the interchange of $x_j$ and $x_{j+1}$. 
Thus, $\psi_\mathrm{AUX}(x_1,\ldots,x_j,x_{j+1},\ldots,x_N,t)$ is antisymmetric 
with respect to the interchange of $x_j$ and $x_{j+1}$, 
which leads to \cite{Gaudin1983,Korepin1993}
\begin{equation}
\psi_\mathrm{AUX}(x_1,\ldots,x_j,x_{j+1},\ldots,x_N,t)|_{x_{j+1}=x_{j}}=0.
\end{equation}
This is fully equivalent to the cusp condition (\ref{interactions}), 
$\hat B_{j+1,j} \psi_B|_{x_{j+1}=x_{j}}=0$ \cite{Gaudin1983,Korepin1993}. 
Thus, the wave function (\ref{ansatz}) obeys constraint 
(\ref{interactions}) by construction. 

In order to exactly describe the dynamics of LL gases, the wave function 
(\ref{ansatz}) should also obey Eq.~(\ref{free}) inside $R_1$. 
From the commutators $[\partial^2/\partial x_j^2,\hat O]=0$ and 
$[i\partial/\partial t,\hat O]=0$ it follows that 
if $\psi_F$ is given by Eq.~(\ref{Slater}) 
and the $\phi_m(x_j,t)$ obey Eq.~(\ref{master}), 
then $\psi_B$ obeys Eq.~(\ref{free}). 
Note that for $c\rightarrow\infty$, one recovers Girardeau's Fermi-Bose 
mapping \cite{Girardeau1960,Girardeau2000}, i.e., $\hat O=1$.

Let us utilize this formalism to describe the dynamics
of a freely expanding LL gas. Suppose that for $t<0$ 
the system is confined by an external potential $V(x)$ 
and is in its ground state, before at $t=0$ the potential is suddenly 
switched off. In order to find the exact form of the initial condition, 
we have to solve the static Schr\" odinger equation for 
the LL gas in the potential $V(x)$. 
By using the above formalism, we express the initial state as 
$\psi_{B0}={\mathcal N}_c \hat O \psi_{F0}$, 
which should (within $R_1$) obey $\sum_j H_j \psi_{B0}= E_{B0} \psi_{B0}$ 
or, equivalently, 
\begin{equation}
\hat O \sum_j H_j \psi_{F0}-
[\hat O , \sum_j H_j] \psi_{F0}
= \hat O E_{B0}\psi_{F0}.
\label{F0gs}
\end{equation}
Here, $E_{B0}$ is the ground-state energy, and 
$H_j=-\partial^2/\partial x_j^2+V(x_j)$ is the SP hamiltonian. 
Eq.~(\ref{F0gs}) shows that, due to the nonvanishing commutator 
$[\hat O , \sum_j H_j]=[\hat O , \sum_j V(x_j)]$, operating with $\hat O$ on 
the fermionic ground state in the trap does not give the bosonic ground state. 
However, for sufficiently 
strong interactions and/or weak and slowly varying potentials, we can 
approximate $[\hat O , \sum_j V(x_j)]\approx 0$. 
In the TG limit, the commutator vanishes identically. 
Thus, for sufficiently strong interactions, the  
ground state is approximated by $\psi_{B0}={\mathcal N}_c \hat O 
\det[\phi_m(x_j,0)]_{m,j=1}^{N}/\sqrt{N!}$, where $\phi_m(x_j,0)$ is the $m$th
eigenstate of the SP hamiltonian.

\begin{figure}
\begin{center}
\includegraphics[width=0.44 \textwidth ]{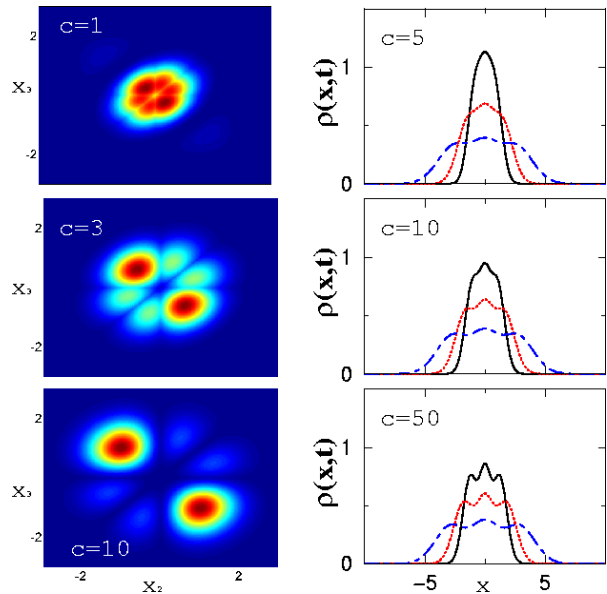}
\caption{ \label{fig}
(color online) The correlation properties of the initial state 
and the evolution of the SP density for various interaction strengths, for
$N=3$ particles.
(Left column) The probability density $|\psi_{B0}(0,x_2,x_3)|^2$ for $c=1,3,10$. 
(Right column) The SP density at $t=0$ (black solid line), $t=0.5$ 
(red dotted line), and $t=1$ (blue dot-dashed line), for $c=5,10,50$.
}
\end{center}
\end{figure}

In what follows we study free expansion from such an 
initial condition, which describes a LL gas with a localized density 
distribution. Even though $\psi_{B0}$ can not be interpreted as 
a ground state for weak interactions, the free expansion from 
$\psi_{B0}$ is calculated exactly for all values of $c$. 
We illustrate the properties of $\psi_{B0}={\mathcal N}_c \hat O 
\det[\phi_m(x_j,0)]_{m,j=1}^{N}/\sqrt{N!}$ for the harmonic potential 
$V(x)=\nu^2 x^2/4$, with $\nu=2$. 
Fig.~\ref{fig} (left column) displays the section $|\psi_{B0}(0,x_2,x_3)|^2$ of the 
probability density, for $N=3$ particles and three values
of $c$. 
We clearly see that, as the interaction strength increases, the 
initial state becomes more correlated. Given that one particle is 
located at zero, for $c=1$ there is a considerable probability 
that the other two particles are to the left or to the right 
of the first one, i.e., their positions are weakly correlated 
with that of the first particle. 
However, for larger $c$, if one particle is at zero, it is more likely 
that the other two particles are on opposite sides of 
the first one, and their distance grows with increasing interaction strength.

When the harmonic potential is turned off, 
the evolution of the SP states $\phi_m(x,t)$ 
is known exactly (see, e.g., Ref.~\cite{Minguzzi2005}): 
$\phi_m(x,t)=\phi_m(x/b(t),0)\exp[ix^2 b'(t)/(4b(t))-iE_m\tau(t)] /\sqrt{b(t)}$,
where $E_m$ is the energy of the $m$th SP eigenstate 
$\phi_m(x,0)$, $b(t)=\sqrt{1+t^2\nu^2}$, and $\tau(t)=\arctan(\nu t)/\nu$.
We can make use of the expression for the ground state of a TG gas 
in harmonic confinement \cite{Girardeau2001} to calculate $\psi_{F0}$.
Employing Eq. \ref{ansatz}, the evolution of the many-body 
wave function $\psi_B$ can then be formally expressed (within $R_1$) as
\begin{eqnarray}
\psi_B={\mathcal N}(c,\nu,N)
b(t)^{-\frac{N^2}{2}}
e^{-i \frac{N^2 \nu}{2} \tau(t)}
\nonumber \\
\hat O
e^{- \frac{\nu-i \nu^2 t}{4} \sum_{j=1}^N [x_j/b(t)]^2 }
\prod_{1\leq i < j \leq N}(x_j-x_i),
\label{exactTD}
\end{eqnarray}
where ${\mathcal N}(c,\nu,N)$ is a normalization constant,
evaluating to $\mathcal{N}=1$ for $c\to\infty$.
The action of the 
operator $\hat O$ yields lengthy expressions already for a few
particles, and particular examples will be given elsewhere. 
The asymptotic form of Eq. (\ref{exactTD}) is given by 
$\lim_{t\rightarrow\infty}\psi_B/\psi_F\propto
\{  
\hat O \prod_{1\leq i < j \leq N}(x_j-x_i)
\}
/\prod_{1\leq i < j \leq N}(x_j-x_i)$.
Although Eq.~(\ref{exactTD}) provides an exact wave function for the 
time-dependent LL gas, it is desirable to calculate the evolution of 
observables such as the SP density $\rho(x,t)=N\int dx_2\ldots dx_N
|\psi_B(x,x_2,\ldots,x_N,t)|^2 $. This task is complicated by the many-fold 
integral.
However, we can find the evolution of $\rho(x,t)$ numerically for small 
numbers of particles. 
Fig. \ref{fig} (right column) displays the evolution of the SP density for three different 
values of $c$. For larger $c$, the initial 
SP density exhibits typical TG-fermionic properties, characterized
by $N$ small separated humps 
\cite{Girardeau2000,Girardeau2000a}. For all values of $c$, 
the SP density acquires such humps during free expansion 
indicating that the system becomes more correlated in time, which is in accord with 
the fact that the LL gas becomes strongly correlated for 
lower densities \cite{Lieb1963}. 
%
%

Even though the employed approach is valid at any interaction strength, 
it is particularly useful for the strongly interacting gas: 
First, the operator $\hat O$ can be hierarchically organized 
into orders $1/c^k$, $\hat O=1+\sum_{k=1}^{N(N-1)/2}c^{-k}\hat O_k$. 
By keeping the terms of order $1/c$, we obtain the first-order 
correction to the TG gas. In this approximation, 
the form of the operator is considerably simplified, 
$\hat O\approx 1+c^{-1}\hat O_1$, where 
$\hat O_1=\sum_{k=1}^{N} (2k-N-1)\partial/\partial x_k$. 
The wave function reads, within $R_1$,
\begin{equation}
\psi_B=\psi_{F}+
\sum_{k=1}^{N} \frac{2k-N-1}{c\sqrt{N!}} \det[A_{mj}^{k}]_{m,j=1}^{N},
\label{wfc}
\end{equation}
where $A_{mj}^{k}=\phi_m(x_j,t)$ for $j\neq k$, and 
$A_{mk}^{k}=\partial\phi_m(x_k,t)/\partial x_k$. 
The numerical calculation 
of the wave function (\ref{wfc}) is not a difficult task 
even for a fairly large numbers of particles. 
Second, in this regime, $[\hat O , \sum_j H_j]=[\hat O , \sum_j V(x_j)]\approx 0$ 
is a reasonable approximation. 
For example, with $\hat O \approx 1+c^{-1}\hat O_1$
and $V(x)=\nu^2 x^2/4$, $[\hat O , \sum_j V(x_j)]$ is of order $\nu^2/4c$,
i.e., for $\nu^2\leq 1/c$ the commutator is of order $1/c^2$ or less. 
Thus, the approach can be used to characterize time-dependent 
and static LL gases in various trapping potentials in the strongly 
correlated regime, but below the TG gas limit.

For completeness, let us briefly discuss time-evolving states $\psi_B$ 
with periodic boundary conditions as in Ref. \cite{Lieb1963}. 
Any time-evolving state $\psi_B$ can be written as a 
superposition of eigenstates $\psi_{\mathrm{LL},\xi}(x_1,\ldots,x_N)$, where 
$\xi$ denotes all quantum numbers necessary to describe 
one eigenstate. LL eigenstates can be written as 
$\psi_{\mathrm{LL},\xi}  = \hat O {\mathcal N}_{\xi} \det[e^{i k_{m}x_j}]_{m,j=1}^{N}$
\cite{Gaudin1983,Korepin1993}. 
If periodic boundary conditions are imposed as in Ref. \cite{Lieb1963}, 
the quasimomenta $k_j$ must obey a set of coupled transcendental 
equations and depend on $c$ \cite{Lieb1963,Sakmann2005}. 
Time-evolving states $\psi_B$ can be written 
as a superposition of LL eigenstates
\begin{equation}
\psi_B =\hat O
\sum_{\xi} {\mathcal N}_{\xi}
b(\xi)\det[e^{i k_{m}x_j}]_{m,j=1}^{N} e^{-iE_{\xi}t},
\end{equation}
where the coefficients $b(\xi)$ are fixed by the initial conditions. 
These coefficient are in practice hard to calculate given the 
initial state due to the many-fold integrations that need to
be performed.

It should be emphasized that the Fermi-Bose transformation employed here 
differs from the fermion-boson duality discussed by Cheon and Shigehara 
\cite{Cheon1999} (see also \cite{Yukalov2005}), because it 
transforms a {\em noninteracting} fermionic wave function into a wave 
function describing LL gas. 
Using Ref.~\cite{Cheon1999} it can be shown that the approach 
used here can also be applied to construct wave functions for a 
time-dependent Fermi gas with finite-strength interactions.

In conclusion, we have constructed exact solutions for the freely expanding LL gas with localized initial density distribution. 
Wave functions are obtained by differentiating a fully antisymmetric 
(fermionic) time-dependent wave function, which obeys the Schr\" odinger 
equation for a free Fermi gas. 
For a number of physically interesting situations (e.g., free expansion), 
by using the operator $\hat O$, the state of the system can be derived from 
$N$ SP time-dependent states, as anticipated in Ref.~\cite{Girardeau2003}.
The construction of LL wave functions for various external 
potentials $V(x)$, and the derivation of correlation functions 
within the employed formalism is the subject of ongoing work.

The authors are grateful to M.~Girardeau for pointing out to them that the formulation 
of the Bethe ansatz as in Eq.~(\ref{ansatz}) was previously stated in Refs.~\cite{Gaudin1983,Korepin1993}. 
H.B. and R.P. acknowledge support by the Croatian Ministry of 
Science (MZO\v S) (Grant No.~119-0000000-1015). 
T.G. acknowledges support by the Deutsche Forschungsgemeinschaft.
This work is also supported by the Croatian-German 
scientific collaboration funded by DAAD and MZO\v S.

\end{document}